\documentclass[twocolumn]{aastex631}

\newcommand{\affgoddard}{Astrophysics Science Division, NASA Goddard Space Flight Center, 8800 Greenbelt Road, Greenbelt, MD 20771, USA}

\newcommand{\affuofa}{Steward Observatory, University of Arizona, 933 North Cherry Avenue, Tucson, AZ 85721, USA}

\begin{document}

\title{JADES: The Star Formation and Dust Attenuation Properties of Galaxies at $\rm{3<z<7}$}

\author[0000-0001-5962-7260]{Charity Woodrum}
\affiliation{\affgoddard}

\author[0000-0003-4702-7561]{Irene Shivaei}
\affiliation{Centro de Astrobiolog\'ia (CAB), CSIC–INTA, Cra. de Ajalvir Km. 4, 28850- Torrej\'on de Ardoz, Madrid, Spain}

\author[0000-0002-7595-121X]{Joris Witstok}
\affiliation{Cosmic Dawn Center (DAWN), Copenhagen, Denmark}
\affiliation{Niels Bohr Institute, University of Copenhagen, Jagtvej 128, DK-2200, Copenhagen, Denmark}

\author[0000-0001-5333-9970]{Aayush Saxena}
\affiliation{Department of Physics, University of Oxford, Denys Wilkinson Building, Keble Road, Oxford OX1 3RH, UK}
\affiliation{Department of Physics and Astronomy, University College London, Gower Street, London WC1E 6BT, UK}

\author[0000-0003-4770-7516]{Charlotte Simmonds}
\affiliation{The Kavli Institute for Cosmology (KICC), University of Cambridge, Madingley Road, Cambridge, CB3 0HA, UK}
\affiliation{Cavendish Laboratory, University of Cambridge, 19 JJ Thomson Avenue, Cambridge, CB3 0HE, UK}

\author[0000-0001-6010-6809]{Jan Scholtz}
\affiliation{The Kavli Institute for Cosmology (KICC), University of Cambridge, Madingley Road, Cambridge, CB3 0HA, UK}
\affiliation{Cavendish Laboratory, University of Cambridge, 19 JJ Thomson Avenue, Cambridge, CB3 0HE, UK}

\author[0000-0003-0883-2226]{Rachana Bhatawdekar}
\affiliation{European Space Agency (ESA), European Space Astronomy Centre (ESAC), Camino Bajo del Castillo s/n, 28692 Villanueva de la Ca\~nada, Madrid, Spain}

\author[0000-0002-8651-9879]{Andrew J.\ Bunker}
\affiliation{Department of Physics, University of Oxford, Denys Wilkinson Building, Keble Road, Oxford OX1 3RH, UK}

\author[0000-0002-6719-380X]{Stefano Carniani}
\affiliation{Scuola Normale Superiore, Piazza dei Cavalieri 7, I-56126 Pisa, Italy}

\author[0000-0003-3458-2275]{St\'ephane Charlot}
\affiliation{Sorbonne Universit\'e, CNRS, UMR 7095, Institut d'Astrophysique de Paris, 98 bis bd Arago, 75014 Paris, France}

\author[0000-0002-2678-2560]{Mirko Curti}
\affiliation{European Southern Observatory, Karl-Schwarzschild-Strasse 2, 85748 Garching, Germany}

\author[0000-0002-9551-0534]{Emma Curtis-Lake}
\affiliation{Centre for Astrophysics Research, Department of Physics, Astronomy and Mathematics, University of Hertfordshire, Hatfield AL10 9AB, UK}

\author[0000-0002-7636-0534]{Jacopo Chevallard}
\affiliation{Department of Physics, University of Oxford, Denys Wilkinson Building, Keble Road, Oxford OX1 3RH, UK}

\author[0000-0003-2388-8172]{Francesco D'Eugenio}
\affiliation{The Kavli Institute for Cosmology (KICC), University of Cambridge, Madingley Road, Cambridge, CB3 0HA, UK}
\affiliation{Cavendish Laboratory, University of Cambridge, 19 JJ Thomson Avenue, Cambridge, CB3 0HE, UK}

\author[0000-0003-4565-8239]{Kevin Hainline}
\affiliation{\affuofa}

\author[0000-0003-4337-6211]{Jakob M.\ Helton}
\affiliation{Department of Astronomy \& Astrophysics, The Pennsylvania State University, University Park, PA 16802, USA}

\author[0000-0002-4985-3819]{Roberto Maiolino}
\affiliation{The Kavli Institute for Cosmology (KICC), University of Cambridge, Madingley Road, Cambridge, CB3 0HA, UK}
\affiliation{Cavendish Laboratory, University of Cambridge, 19 JJ Thomson Avenue, Cambridge, CB3 0HE, UK}

\author[0000-0002-0362-5941]{Michele Perna}
\affiliation{Centro de Astrobiolog\'ia (CAB), CSIC–INTA, Cra. de Ajalvir Km.~4, 28850- Torrej\'on de Ardoz, Madrid, Spain}

\author[0000-0002-5104-8245]{Pierluigi Rinaldi}
\affiliation{Space Telescope Science Institute, 3700 San Martin Drive, Baltimore, Maryland 21218, USA}

\author[0000-0002-4271-0364]{Brant Robertson}
\affiliation{Department of Astronomy and Astrophysics, University of California, Santa Cruz, 1156 High Street, Santa Cruz, CA 95064, USA}

\author[0000-0002-4772-7878]{Amber Straughn}
\affiliation{\affgoddard}

\author[0000-0001-6561-9443]{Yang Sun}
\affiliation{\affuofa}

\author[0000-0002-8224-4505]{Sandro Tacchella}
\affiliation{The Kavli Institute for Cosmology (KICC), University of Cambridge, Madingley Road, Cambridge, CB3 0HA, UK}
\affiliation{Cavendish Laboratory, University of Cambridge, 19 JJ Thomson Avenue, Cambridge, CB3 0HE, UK}

\author[0000-0003-2919-7495]{Christina C.\ Williams}
\affiliation{NSF National Optical-Infrared Astronomy Research Laboratory, 950 North Cherry Avenue, Tucson, AZ 85719, USA}

\author[0000-0002-4201-7367]{Chris Willott}\affiliation{NRC Herzberg, 5071 West Saanich Rd, Victoria, BC V9E 2E7, Canada}

\author[0000-0003-3307-7525]{Yongda Zhu}
\affiliation{\affuofa}

\begin{abstract}

We present the star formation and dust attenuation properties for a sample of 602 galaxies at redshifts $\rm{3<z<7}$, as part of the JADES survey. 
Our analysis is based on measurements of the $\rm{H}\alpha/\rm{H}\beta$ Balmer Decrement using medium resolution (R$\sim$1000) spectroscopic observations with the JWST/NIRSpec Micro-Shutter Assembly. Stellar masses and star formation rates (SFRs) are inferred with \texttt{Prospector} using deep multi-band imaging.
We utilize the Balmer decrement to measure dust-corrected H$\alpha$-based SFRs, taking into account the subsolar metallicities observed in galaxies at high redshift.
We confirm, with our large sample size, that the correlation between the Balmer decrement and stellar mass is already established out to $z\sim7$. We find that the relation between the Balmer decrement and stellar mass does not significantly evolve from the local universe to $z\sim7$.
We investigate the UV slope as a function of the Balmer optical depth and find that the best-fit correlation for our high redshift sample is sSFR dependent and significantly different at high redshift when compared to galaxies at $z\approx 0$ and $z \approx 2$. For the highest sSFR galaxies in our sample, there is no significant correlation between the UV slope and Balmer optical depth. This is evidence that the UV slope should be used with great caution to correct for dust in high redshift galaxies.

\end{abstract}

\NewPageAfterKeywords
\section{Introduction} \label{sec:intro}
Dust is one of the fundamental components of a galaxy and contains information about a galaxy’s star formation and chemical enrichment history \citep{Madau2014}. The Balmer recombination lines, H$\alpha$ and H$\beta$, are often considered the ``gold standard" probes of both star formation and dust attenuation in galaxies \citep{Calzetti2000, ForsterSchreiber2011, Kennicutt2012, Price2014, Shivaei2015, Shivaei2016, Theios2019, Matharu2023}. Dust-corrected H$\alpha$ fluxes can be used to infer instantaneous star formation rates (SFRs). The flux ratio of $\rm{H}\alpha/\rm{H}\beta$, called the Balmer decrement, can be used to measure the nebular dust attenuation towards star-forming regions. Studying these parameters as a function of other galaxy properties can provide clues about galaxy populations and how they evolve over time. 

The significant correlation between stellar mass and dust attenuation has been well studied at low redshifts $z\leq$0.5. Studies have shown that with increasing stellar mass, galaxies in the local universe show increasing dust attenuation, as measured by the Balmer decrement \citep{Groves2012}. Further studies, using results from HST and ground-based NIR surveys, found that this trend was already established at redshifts $z\approx1-3$ \citep{Dominguez2013, Reddy2015, Nelson2016, Whitaker2017, Lorenz2023, Matharu2023, Shivaei2020, Maheson2024}. This implies that more massive galaxies have produced more dust and/or are able to retain more dust, from the local universe to cosmic noon.

Until recently, it has been impossible to measure the Balmer decrement for populations of galaxies at $z\gtrsim 3$. Past this redshift, H$\alpha$ is redshifted beyond $2.6\mu m$, shifting it outside the observing window of ground-based near-infrared instruments. With the launch of JWST \citep{Gardner2023, McElwain2023, Rigby2023}, both H$\alpha$ and H$\beta$ can be measured out to redshift $z\sim7$ with NIRSpec. Early results using JWST have shown that the correlation between the Balmer decrement and M$_*$ is already in place at these high redshifts \citep{Shapley2023, Sandles2024}.

In this work, we utilize medium resolution spectroscopy and deep multi-band photometry from the JWST Advanced Deep Extragalactic Survey \citep[JADES;][]{Eisenstein2023a, Rieke2023, Bunker2024} to investigate the star formation and dust attenuation properties of galaxies at $3<z<7$. In Section \ref{sec:data}, we describe our data, sample selection, and measurements. We present our results in Section \ref{sec:results}, our discussion in Section \ref{sec:discussion}, and a summary in Section \ref{sec:summary}.

\section{Data and Sample Selection} \label{sec:data}
\subsection{JADES}\label{sec:jades}
We use the full JADES \citep{Bunker2024, Eisenstein2023a, D'Eugenio2025, Scholtz2025} photometric and spectroscopic set in the GOODS-N and GOODS-S regions, including the JADES Origins Field \citep[JOF;][]{Eisenstein2023b}, the NIRCam Deep imaging \citep{Rieke2023}, and when available, from the JWST Extragalactic Medium-band Survey \citep[JEMS;][]{Williams2023} and the First Reionization Epoch Spectroscopic Complete Survey \citep[FRESCO;][]{Oesch2023}.

The spectroscopic observations used in this paper were observed with the JWST using the micro-shutter assembly (MSA) of NIRSpec \citep{Jakobsen2022, Ferruit2022, Boker2023}. In order to resolve H$\alpha$ from the [\ion{N}{2}] lines, we utilize observations from the medium resolution gratings (G140M/F070LP, G235M/F170LP, and G395M/F290LP) with spectral resolution $R\sim 1000$ across the full spectral range of NIRSpec. For more information, see \citet{Bunker2024, D'Eugenio2025, Scholtz2025}.

We select galaxies that have medium resolution spectroscopic observations as part of JADES, described in Section \ref{sec:jades}. We require the H$\alpha$ line flux S/N$\geq5$ and H$\beta$ line flux S/N$\geq 3$. We restricted our redshift range to $3\leq z \leq 7$, noting that H$\alpha$ is redshifted out of the spectral range of NIRSpec at $z\sim$7. Because the stellar masses used in this work are inferred from photometry, we required galaxies to have JADES NIRCam observations.

To exclude galaxies with AGN activity, we utilize the collected samples of AGN in GOODS-N and GOODS-S \citep{Hainline2024}. We remove the sample of broad-line, Type I AGN spectroscopically selected from JWST/NIRSpec observations of JADES targets from \cite{Juodzbalis2025} and \cite{Maiolino2024} and the Type II AGN identified in \citet{Scholtz2023}. After cross-matching with our sample, we remove a total of twenty-five Type I AGN and thirteen Type II AGN. 

The UV slopes are taken from \citet{Saxena2024} and were measured using $R\sim100$ JWST PRISM/CLEAR spectra by fitting a power law to the rest-frame wavelength range 1340-2700\AA, for more details see \citet{Saxena2024}.

\subsection{Sloan Digital Sky Survey Comparison Sample}
For our local universe comparison sample, we utilize the Sloan Digital Sky Survey (SDSS) Data Release 7 \citep[DR7;][]{Abazajian2009}. We use the MPA-JHU catalog of measurements for DR7\footnote{Available at \url{http://www.mpa-garching.mpg.de/SDSS/DR7/}}. After removing the duplicates in the catalog, we restrict the redshift range to $0.04\leq z \leq 0.10$ to minimize aperture effects, following \citet{Shapley2022}. Futhermore, we require a S/N of $>5$ for H$\beta$, H$\alpha$, and [\ion{N}{2}]$\lambda$6584, and a S/N of $>3$ for [\ion{O}{3}]$\lambda$5007. We exclude galaxies that meet the optical emission line diagnostic for AGN based on the \citet{Kauffmann2003} criteria. This renders a local universe sample size of $\sim 105,000$ galaxies.

\begin{figure*}[!htbp]
\includegraphics[width=\textwidth]{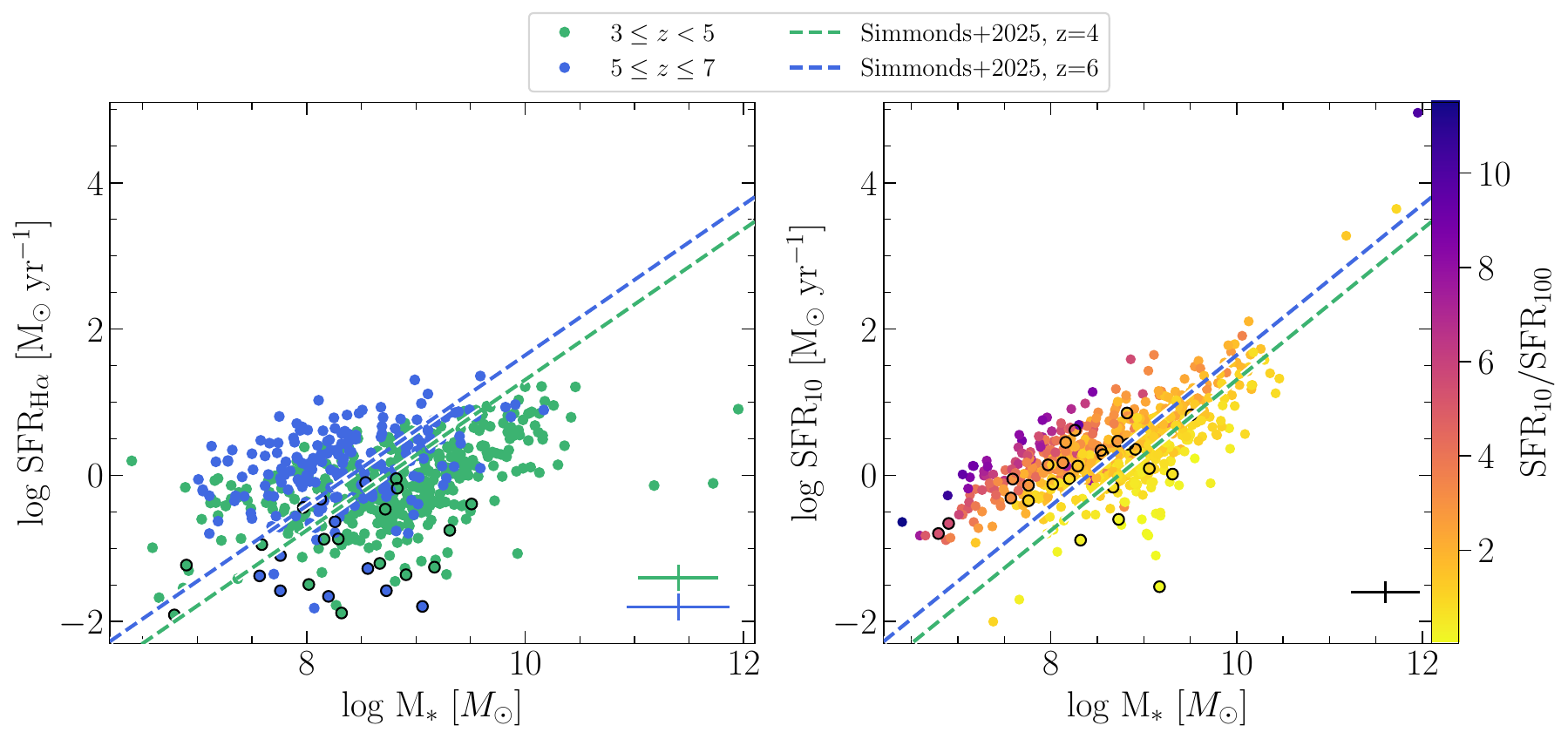}
\caption{The left panel shows the dust-corrected $\mathrm{SFR}_{\rm H\alpha}$ vs. the stellar mass for our sample of galaxies. The solid lines show the best-fit relation from \cite{Simmonds2025} for the median redshifts in the two different bins, at $z=4$ and $z=6$. Galaxies with low Balmer decrement measurements, $H\alpha/H\beta<2.75$ within their $3\sigma$ uncertainties, are shown with black marker outlines. Typical uncertainties are shown in the lower right corner. The right panel shows the SFR averaged over the last 10 Myr measured from the SED fitting vs. the stellar mass. The points are color-coded by their burstiness, defined as the fraction of the SFR average over 10 Myr and the the SFR sustained over 100 Myr. 92\% of galaxies with burstiness values $>5$ have low stellar masses $M_*< 10^{8.5} \rm{M}_{\odot}$. \label{fig:MS}}
\end{figure*}

\subsection{Analysis}
\subsubsection{Line Measurements}
Continuum and line emission for the JADES spectra were modeled using \texttt{QubeSpec’s} fitting module. The continuum was fitted as a power law and the emission lines were fitted using a single Gaussian. The model parameters were estimated using the Bayesian fitting code \texttt{QubeSpec}, which utilizes the Markov Chain Monte Carlo (MCMC) integrator \texttt{emcee} \citep{Foreman-Mackey2013}. The fluxes and uncertainties are calculated as the median and standard deviation of the MCMC chains (after discarding the burn-in chains). For more information, including details of the data reduction process, see \citet{Scholtz2025}. 

\subsubsection{SED Fitting}
The stellar population properties used in this work, including SFR$_{\rm{SED}}$ and the stellar mass $M_*$, were inferred using the SED fitting code \texttt{Prospector} \citep{Johnson2021}, which uses the Flexible Stellar Population Synthesis (FSPS) code \citep{Conroy2009} via Python-FSPS \citep{Foreman-Mackey2014}. We used JWST NIRCam photometry from JADES including the following bands: F909W, F115W, F150W, F162M, F200W, F250M, F277W, F300M, F335M, F356W, F410M, and F444W. JEMS medium band photometry was used when available including the following 5 photometric bands: F182M, F210M, F430M, F460M, and F480M. Ancillary photometry from HST was fit simultaneously including the ACS bands: F435W, F606W, F775W, F814W, F850LP, and the HST IR bands: F105W, F125W, F140W, and F160W \citep{Whitaker2019}. The details of the fitting procedure can be found in \citet{Simmonds2024}, but are summarized briefly here. The \citet{Chabrier2003} initial mass function (IMF) is adopted, with mass cutoffs of 0.1 and 100$M_{\odot}$. Dust is modeled using a two component dust model \citep{Charlot2000} and a variable dust index \citep{Kriek2013}. Nebular emission is based on \texttt{CLOUDY} models using MESA Isochrones \& Stellar Tracks \citep[MIST;][]{Choi2016, Dotter2016} and the MILES stellar library \citep{Vazdekis2015}. The intergalactic medium (IGM) absorption is modeled after \citet{Madau1995} and the scaling of the IGM attenuation is a free parameter. For the star formation history (SFH), we use a nonparametric model \citep{Leja2019} with a bursty-continuity prior \citep{Tacchella2022b} with eight distinct SFR bins. Throughout this paper, SFR$_{\rm{SED}}$ refers to the most recent SFR averaged over the last 10 Myr.

\subsubsection{SFR Measurements} \label{sec:SFR}
In addition to SFR$_{\rm{SED}}$, we also measure SFR$_{\rm{H}\alpha}$ using dust-corrected H$\alpha$ luminosities. Using the observed Balmer decrement, H$\alpha/\rm{H}\beta$, we infer the nebular attenuation, E(B-V)$_{\rm{neb}}$. The \citet{Cardelli1989} dust extinction curve has been shown to accurately represent the nebular attenuation in star-forming galaxies at $z\sim2$ \citep{Reddy2020}. We therefore assume the \citet{Cardelli1989} law to infer the nebular attenuation as E(B-V)$_{\rm{neb}} = 2.33\ \log\ (\frac{\rm{H}\alpha/\rm{H}\beta}{2.86})$ \citep{Shivaei2020}. To convert from dust-corrected H$\alpha$ luminosities to SFRs, as done in \citet{Shapley2023} and following \citet{Du2018}, we adopt differing conversion factors based on the individual redshift and mass of each galaxy, motivated by the evolving mass-metallicity relation \citep[e.g.,][]{Sanders2021}. For galaxies at $z<3.4$ and stellar masses $\gtrsim 10^{10.66} \rm{M}_{\odot}$ we adopt the conversion factor of $10^{-41.37} (\rm{M_{\odot}\ yr^{-1})}/(erg\ s^{-1})$, which is derived from $Z_* = 0.02$ BPASS population synthesis models, includes the effects of stellar binaries, and assumes an upper limit on the IMF mass of 100M$_{\odot}$. Only one galaxy meets this criteria. At redshifts $z>3.4$ for all stellar masses, we adopt the conversion factor of $10^{-41.67} (\rm{M_{\odot}\ yr^{-1})}/(erg\ s^{-1})$, which is derived from $Z_* = 0.001$ BPASS population synthesis models, includes the effects of stellar binaries, and assumes an upper limit on the IMF mass of 100M$_{\odot}$ \citep{Reddy2022}. Importantly, these conversion factor account for the subsolar metallicities common in galaxies at high redshift \citep[e.g.,][]{Cullen2019, Sanders2021}.

\section{Results}\label{sec:results}
\subsection{Star Forming Main Sequence}
One of the fundamental indicators of the evolutionary state of star-forming galaxies is its location in the SFR-M$_*$ plane, often called the star-forming main sequence \citep[SFMS;][]{Daddi2007, Noeske2007, Salim2007, Karim2011, Rodighiero2011, Whitaker2012, Speagle2014, Whitaker2014, Renzini2015, Schreiber2015, Tomczak2016, Leslie2020, Leja2022, Rinaldi2022, Rinaldi2025}. The existence of the SFMS suggests that similar physical mechanisms could be responsible for the growth of low- and high-mass galaxies alike \citep{Noeske2007}.  Galaxies are thought to grow in mass along this sequence, with galaxies significantly above the SFMS being in a starburst phase \citep{Muxlow2006}, while galaxies significantly below the SFMS are considered quiescent or in the process of quenching \citep{Tacchella2016}. The intrinsic scatter of the SFMS may indicate that galaxies have differing star formation histories at a given stellar mass \citep{Matthee2019}. Therefore, to understand the evolutionary state of galaxies, it is important to measure robust SFRs.

Some of the most robust tracers of SFRs are the hydrogen recombination lines, specifically, H$\alpha$ traces the SFR over the past 10 Myr \citep{Kennicutt2012, FloresVelazquez2021, Tacchella2022, McClymont2025}. In Section \ref{sec:SFR}, we describe our methodology for inferring SFRs from dust-corrected H$\alpha$ fluxes, which also accounts for the low-metalliticies that are typical in galaxies at the redshifts of our sample.

In figure \ref{fig:MS}, we compare the distribution of SFR versus mass for our sample to the parameterization of the SFMS as a function of redshift from \citet{Simmonds2025}. The majority of galaxies at stellar masses $\lesssim 10^{8.5} \rm{M}_{\odot}$ lie above the \citet{Simmonds2025} relation. Because we selected galaxies with strongly detected emission lines, our sample consists of low-mass galaxies that are in a bursty phase of star formation, while excluding the low-mass galaxies with lower SFRs. Therefore, these higher SFR values in the lower-mass galaxies are due to selection effects \citep{Rinaldi2022, Navarro-Carrera2024, Rinaldi2025}. Low-mass galaxies at high redshift tend to have higher scatter in their SFRs from bursty star formation histories, as predicted in simulations \citep[e.g.,][]{Dayal2013, Kimm2015, Ceverino2018, Faucher-Giguere2018, Ma2018, Rosdahl2018, Barrow2020, Tacchella2020, Furlanetto2022, Dome2024} and found in observations \citep{Looser2023, Asada2024, Boyett2024, Endsley2024, Looser2024}. In the right panel of Figure \ref{fig:MS}, the points are color-coded based on the ratio of their recent SFR over 10 Myr and their sustained SFR over 100 Myr, $\rm{SFR_{10,SED}/SFR_{100,SED}}$. This metric is a measure of the burstiness of star formation, with higher values indicating more bursty SFHs. The percent of galaxies with a burstiness value $\rm{SFR_{10,SED}/SFR_{100,SED}}>5$ and stellar masses $\lesssim 10^{8.5} \rm{M}_{\odot}$ is 94\%. This confirms that the lower mass galaxies in our sample have burstier SFHs.

\begin{figure}[!htbp]
\includegraphics[width=0.5\textwidth]{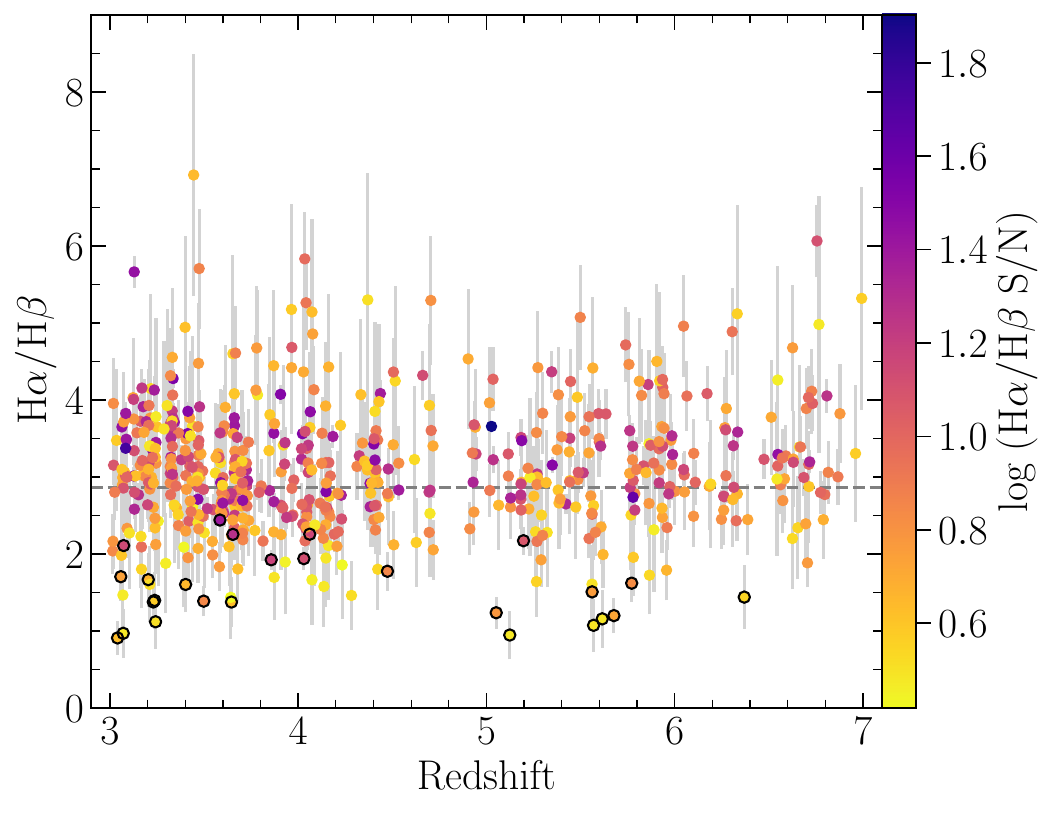}
\caption{Balmer Decrement, H$\alpha/\rm{H}\beta$, vs. the spectroscopic redshift for our sample of 602 galaxies. The points are color-coded based on their H$\alpha/\rm{H}\beta$ S/N. Individual emission lines have an S/N of H$\alpha>5$ and H$\beta>3$. The horizontal dashed line indicates a value of 2.86 for the Balmer Decrement, which is the Case B limit. A total of twenty-six galaxies fall below 2.75 within their 3$\sigma$ uncertainties and are indicated with black marker outlines. \label{fig:bd_redshift_snr}}
\end{figure}

\begin{figure*}[!htbp]
\includegraphics[width=\textwidth]{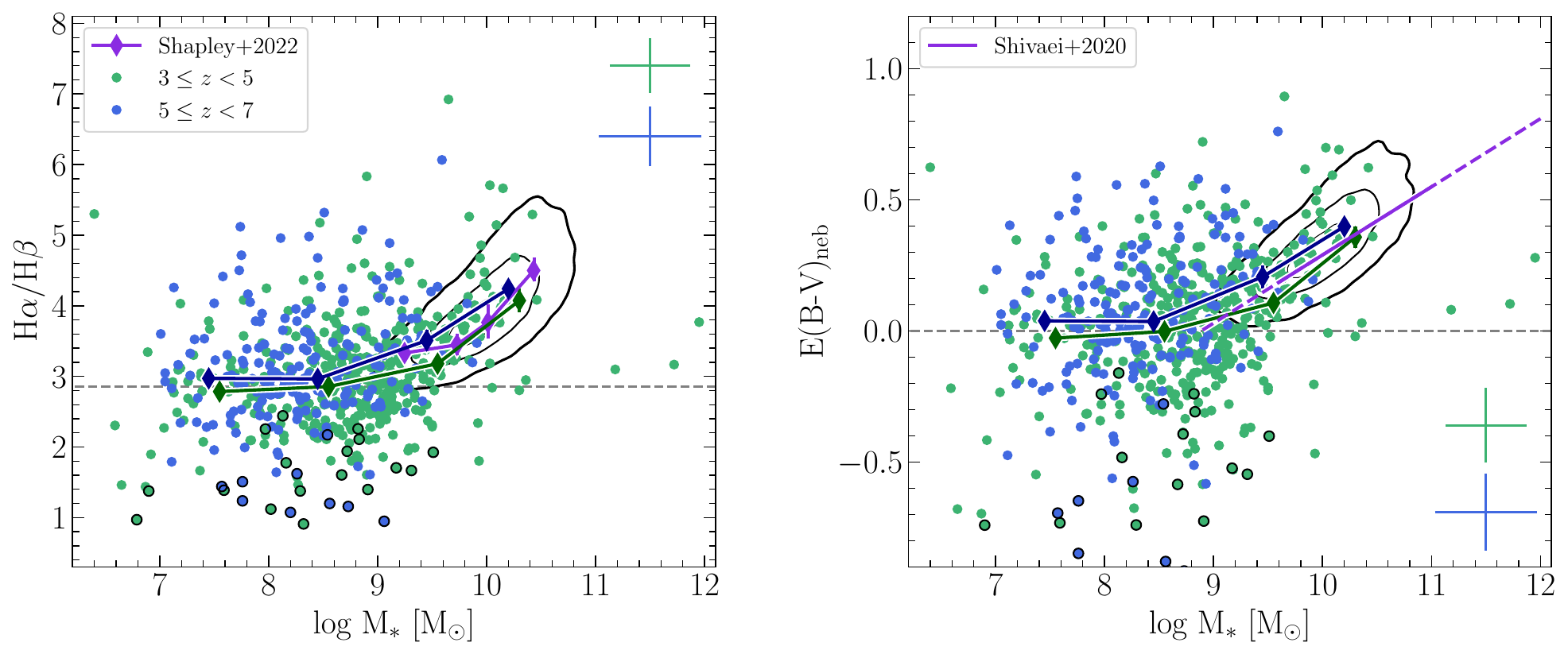}
\caption{The left panel shows the Balmer Decrement, H$\alpha/\rm{H}\beta$, as a function of M$_*$, while the right panel shows the nebular reddening, E(B - V)$_{\rm{neb}}$, as a function of M$_*$. Galaxies in the redshift range $3\leq z<5$ are shown in green, and those in the redshift range $5\leq z<7$ are shown in blue, with their corresponding typical uncertainties in the upper right corner. The green and blue lines represent the medians in four stellar mass bins, with the error on the median shown as errorbars. The black contours show the 16th, 50th, and 84th percentiles for local SDSS galaxies. In the left panel, the purple diamonds and lines show the running median for $z\sim2.3$ star-forming galaxies in the MOSDEF survey from \cite{Shapley2022}. In the right panel, the purple line shows the linear regression fit at z=$1.4-2.6$ from \cite{Shivaei2020}. The dashed line is shown to indicate that the sample from \cite{Shivaei2020} is not complete below a stellar mass of $\sim 10^{9.5}\ M_{\odot}$. \label{fig:bd_ebv_mass}}
\end{figure*}

\subsection{Dust Attenuation}\label{sec:dustattenuation}
In this section we present the dust attenuation properties of the galaxies in our sample. In Figure \ref{fig:bd_redshift_snr}, we show the Balmer decrement, H$\alpha/\rm{H}\beta$, as a function of spectroscopic redshift for individual galaxies. The circles are color-coded based on their H$\alpha/\rm{H}\beta$ S/N. Individual emission lines have S/N$\geq$5 for H$\alpha$ and S/N$\geq3$ for H$\beta$. The unattenuated, or intrinsic, Balmer decrement value for Case B recombination with temperature $\rm{T_e} = 10\ 000$K and density $\rm{n_e} = 100 \rm{cm}^{-3}$ is 2.86 \citep{Osterbrock2006} and is indicated in the plot by a dashed horizontal line. The Balmer decrement can vary with differing temperatures and electron densities. For example, using temperatures and electron densities that are typical in star-forming regions with the ranges $5,000< \rm{T_e} <30,000$K and $10< \rm{n_e} < 500 \rm{cm}^{-3}$ changes the ratio by 5-6 percent \citep{Sandles2024}. Low-mass galaxies at higher redshifts may have higher gas temperatures \citep{Schaerer2022, Curti2023} and higher electron densities \citep{Reddy2023b}. When assuming $\rm{T_e}=20,000K$ and $\rm{n_e}=300 \rm{cm}^{-3}$, the intrinsic Balmer decrement decreases to 2.75.

Some galaxies fall below the intrinsic H$\alpha /\rm{H}\beta$ ratio. We find a total of twenty-six galaxies that fall below 2.75 within their $3\sigma$ uncertainties. We inspect the emission line fits to the spectra for these galaxies and find that they are well-fit by the model. To determine if the low Balmer decrement values may be due to a nearby galaxy being observed in the same slit, we visually inspect the slit positions. We find that none of the observations were contaminated by a nearby companion in the slits. Therefore, it is possible that their Balmer Decrement values are genuinely low. The twenty-six low Balmer decrement galaxies are indicated on all plots with black marker outlines. 

These genuinely low Balmer decrement values cannot be explained by dust attenuation and imply that Case B recombination is not valid for these galaxies. Recent studies have investigated potential explanations for anomalous Balmer emitters. \citet{Yanagisawa2024} and \citet{Scarlata2024} suggest that galaxies with genuinely low Balmer decrements are not optically thin to Balmer lines. This can lead to low Balmer decrement values if an optically thick shell of gas surrounds the emission nebulae \citep{Yanagisawa2024}. Another possible explanation, favored by \citet{McClymont2024}, is density-bounded nebulae, which are nebulae where ionising photons can escape from a cloud of gas. In this scenario, the Balmer line ratios vary with differing depths in a cloud.

The Pearson's rank correlation for the Balmer decrement as a function of redshift is r$=0.09$ with a p-value of 0.04. Therefore, we do not find a statistically significant correlation between the Balmer decrement and redshift at $3<z<7$, suggesting no significant evolution across this redshift range. To further explore the redshift evolution of dust attenuation in our sample, next we analyze the dust content of galaxies as a function of stellar mass, in two redshift bins.

In the left panel of Figure \ref{fig:bd_ebv_mass}, we show the Balmer decrement as a function of stellar mass, in two different redshift bins, $3\leq z<5$ and $5\leq z<7$. For this relation, we find a statistically significant Pearson's rank correlation with r=0.29 and p$\approx 10^{-13}$. This implies that the correlation between the Balmer decrement and stellar mass is in place out to z=3-7, as found previously with smaller sample sizes \citep{Shapley2023, Sandles2024}. In the right panel we show the nebular reddening derived from the Balmer decrement, E(B-V)$_{\rm{neb}}$, as a function of stellar mass compared with the relation found in \citet{Shivaei2020}. The medians of the Balmer decrement as a function of stellar mass for the galaxies in our separate redshift bins are consistent with each other across the entire stellar mass range within the $1\sigma$ uncertainties. Therefore, none of the stellar mass bins show significant evolution in the dust attenuation between the two redshift bins.

Next we compare the galaxies in our sample with those at lower redshift. The Balmer decrement as a function of stellar mass for local SDSS galaxies is shown with the 16th, 50th, and 84th percentiles indicated as black contours. The $z\sim 2.3$ running medians are shown as purple diamonds \citep{Shapley2022} in the left panel and as a linear relation in the right panel \citep{Shivaei2020}. The galaxies in our sample at stellar masses $\gtrsim 10^{9}\ \rm{M}_{\odot}$ have consistent Balmer decrement and nebular reddening values with galaxies in the MOSDEF survey at $z\sim 2.3$ \citep{Shivaei2020, Shapley2022} and local galaxies at $z\sim 0$ from SDSS, within their $1\sigma$ uncertainties, consistent with \citet{Shapley2023, Maheson2025}. This suggests that, at a given mass $\gtrsim 10^{9}\ M_{\odot}$, the dust attenuation of the galaxies in our sample does not evolve with redshift. 

\begin{figure*}[!htbp]
\includegraphics[width=\textwidth]{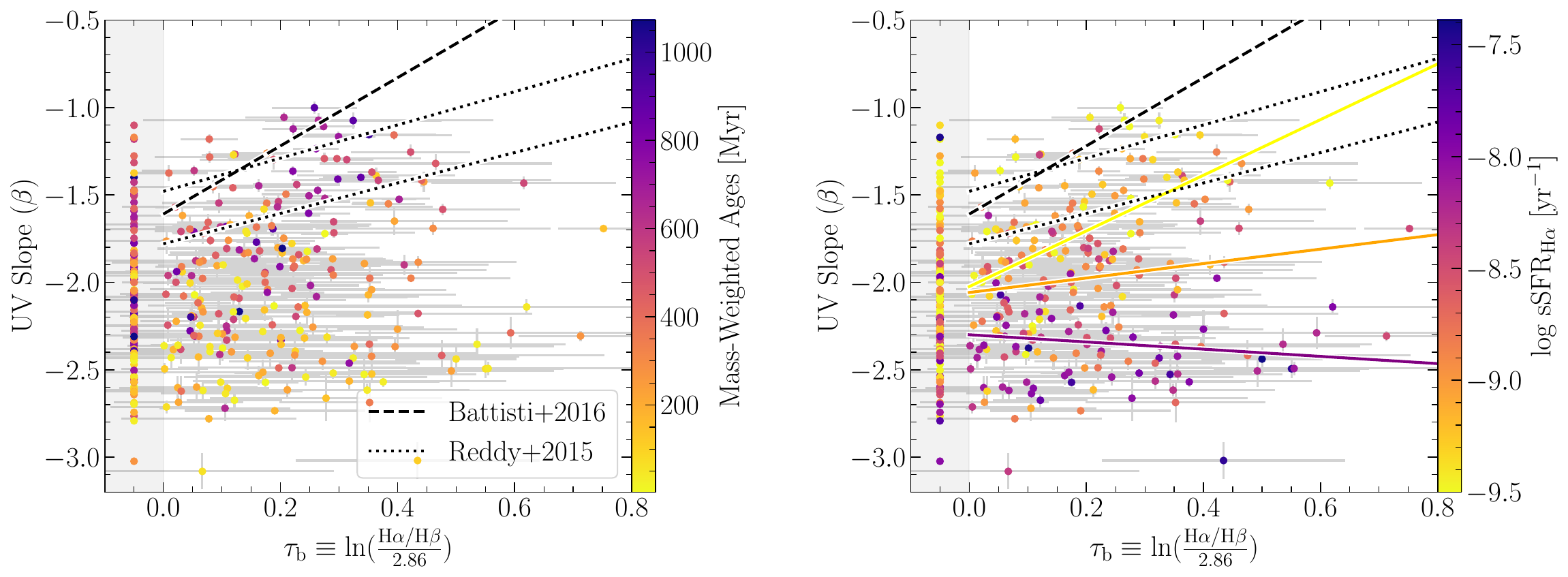}
\caption{UV slope vs. Balmer optical depth. The relation from \citet{Battisti2016} at z=0 is shown as a dashed line while the relations from \citet{Reddy2015} at $z\sim 2$ are shown as dotted lines. \citet{Reddy2015} divided their sample based on sSFR values and found differing relations. The top dotted line shows the relation for galaxies with $-9.60< \log(\mathrm{sSFR/yr^{-1}})< -8.84$ and the bottom line for galaxies with $-8.84< \log(\mathrm{sSFR/yr^{-1}})< -8.00$. We separate our sample in to three bins based on sSFR. The yellow line shows the relation found for galaxies with $-11.8< \log(\mathrm{sSFR/yr^{-1}})< -9$, the orange line shows the relation for galaxies with $-9.00< \log(\mathrm{sSFR/yr^{-1}})< -8.00$, and the purple line shows the relation for galaxies with $-8.00< \log(\mathrm{sSFR/yr^{-1}})< -6.9$. \label{fig:beta_tau}}
\end{figure*}

\section{Discussion}\label{sec:discussion}
Our sample provides further evidence that the correlation between the Balmer decrement and stellar mass does not significantly evolve from the local universe out to $z\sim$7 \citep{Whitaker2017, Shapley2022, Shapley2023, Sandles2024}. Another diagnostic used to infer the dust attenuation properties of star-forming galaxies is the UV spectral slope $\beta$ \citep[e.g.,][]{Nandra2002, Reddy2004, Reddy2006, Reddy2012, Daddi2007, Buat2009, Pannella2009, Magdis2010}. Although the UV slope varies with properties such as metallicities, stellar ages, and star formation histories, it has been shown to be significantly more sensitive to dust attenuation \citep{Calzetti1994, Reddy2015, Wilkins2016, Tacchella2022b}. Therefore, there is an observed correlation between the UV slope and Balmer optical depth.

In Figure \ref{fig:beta_tau}, we show the UV slope \citep[as measured from NIRSpec spectra in][]{Saxena2024} as a function of the Balmer optical depth ($\tau_b$), defined as $\tau_{b}\equiv \ln(\frac{\rm{H}\alpha/\rm{H}\beta}{2.86})$ \citep{Calzetti1994}. We only include galaxies with UV slope measurements with uncertainties that are $<5\%$, rendering a sample size of 389 galaxies for this comparison. The points in the left panel are color-coded based on their mass-weighted ages, which were measured as their half-mass assembly time; the points in the right panel are color-coded based on their specific star formation rates (sSFR$\rm{=M_*/SFR\ yr^{-1}}$).

\begin{figure*}[!htbp]
\includegraphics[width=\textwidth]{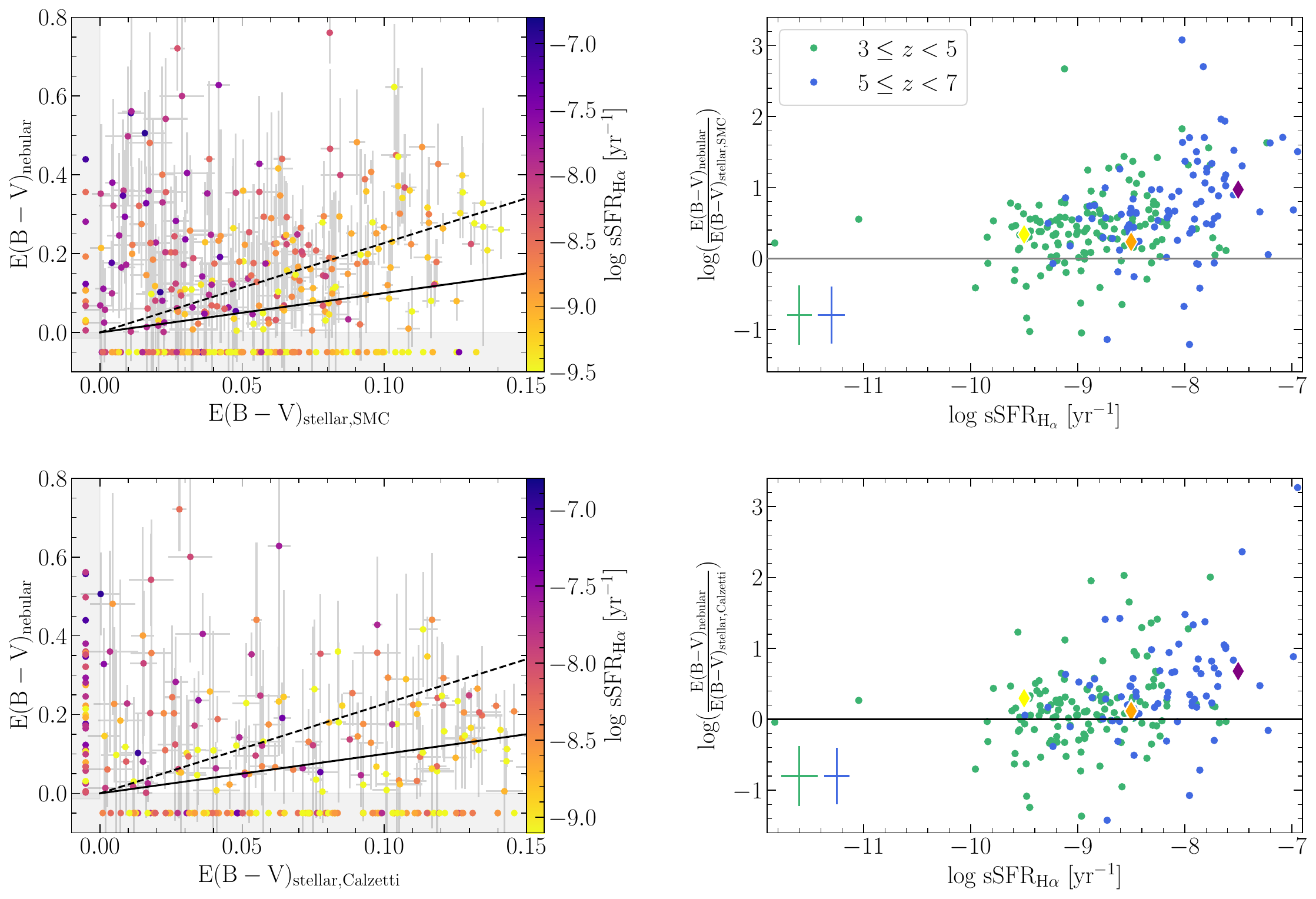}
\caption{The nebular reddening, $\mathrm{E(B-V)_{nebular}}$ as measured by the Balmer decrement, vs. the stellar reddening, $\mathrm{E(B-V)_{stellar}}$ as measured by the UV slope. In the left panels, points are color-coded based on their sSFR. The solid line indicates the one-to-one relation and the dashed line corresponds to the local relation from \citet{Calzetti2000}. Areas where $\mathrm{E(B-V)<0}$ are shaded in gray. The right panels show the reddening ratio, $\mathrm{E(B-V)_{nebular}/E(B-V)_{stellar}}$, as a function of the sSFR. Typical uncertainties are shown in the bottom left corners. The medians are shown as diamonds for three differing sSFR bins. The bootstrapped uncertainties on the median are smaller than the marker size.  \label{fig:ebv_neb_stell}}
\end{figure*}

At a given Balmer optical depth, there is large scatter in the UV slope. \citet{Reddy2015} showed that the scatter for their sample at $z\sim2$ is largely due to variations in sSFRs. They found that at a given Balmer optical depth, galaxies with lower sSFRs have on average redder UV slopes, consistent with our results, see Figure \ref{fig:beta_tau}. We separate our sample into three bins depending on the sSFR of the galaxy, $-11.8<\rm{log(sSFR/yr^{-1})}\leq -9$ and $-9<\rm{log(sSFR/yr^{-1})}\leq -8$ and $-8< \rm{log(sSFR/yr^{-1})}\leq -6.9$. We derive the best-fit relation for $\beta$ as a function of $\tau_b$, only for galaxies with $\mathrm{\tau_b > 0}$, for the three differing sSFR bins. The best-fit relations are shown in Figure \ref{fig:beta_tau} as yellow, orange, and purple lines. We find that the best-fit relations for our sample at high redshift vary significantly between the differing sSFR bins. The differences in the vertical offsets between these relations could be due to differing average stellar population ages, where higher sSFR galaxies have more young stars and bluer stellar continuum, as compared to lower sSFR galaxies. This is consistent with the left panel of Figure \ref{fig:beta_tau}, which is color-coded by the mass-weighted ages, and shows that the higher sSFR galaxies indeed have younger stellar ages. The slopes of the best-fit relations vary between sSFR bins such that galaxies with lower sSFRs have a steeper slope, indicating a stronger correlation between $\beta$ and $\tau_b$. The highest sSFR bin shows no significant correlation between $\beta$ and $\tau_b$ with a Pearson's rank coefficent of $\rho=-0.15$ and a p-value of 0.29. We therefore emphasize that using UV slopes as an indication of the amount of dust in high redshift galaxies is highly dependent on sSFR and age. Galaxies in the highest sSFR bin for our sample have a range of nebular dust attenuation, as measured by $\tau_b$, but on average low UV continuum dust attenuation, as measured by $\beta$.

We also show the best-fit $\beta$ vs. $\tau_b$ relations for galaxies in the local universe \citep{Battisti2016} and at cosmic noon \citep{Reddy2015}. The cosmic noon sample includes galaxies with sSFRs in the range $-9.6<\rm{log(sSFR/yr^{-1})}<-8.00$ and the local sample includes galaxies with sSFRs in the range $-10.5\lesssim \rm{log(sSFR/yr^{-1})} \lesssim-8.9$. As discussed in \citet{Reddy2015}, the differing relations indicate that the correlation between UV stellar continuum and nebular dust attenuation is sSFR dependent and also differs with redshift. \citet{Battisti2016} suggest that this may indicate evolution in the dust with redshift, such as differing geometry, chemical composition, absorption/scattering cross sections, or size distributions. Evolution of the intrinsic UV slope ($\mathrm{\beta_0}$) with redshift may also be playing a role in the differing relations between $\beta$ vs. $\tau_b$. If the intrinsic UV slope is bluer at high redshifts because galaxies have stellar populations with lower metallicities \citep[e.g.,][]{Reddy2018}, then the relations in the right panel of Figure \ref{fig:beta_tau} will move vertically, because their zero-point is changing.

Next, we compare the nebular reddening, $\mathrm{E(B-V)_{nebular}}$ as measured by the Balmer decrement, with the stellar reddening, $\mathrm{E(B-V)_{stellar}}$ as measured by the UV slope. E(B-V) is a measurement of the difference in attenuation between the B and V bands. The nebular reddening is measured from emission lines and shows the dust column density towards ionized nebulae, while the stellar reddening is measured from the continuum emission and shows the dust column density of the stellar continuum. Many studies have shown that the nebular reddening is higher than the stellar reddening in star-forming galaxies \citep{Calzetti2000, Battisti2016}. This can be explained by a two-component dust model, where all stars in the galaxy are reddened by diffuse ISM dust, while star-forming regions are additionally reddened by their birth-could dust content \citep{Calzetti2000, Charlot2000}. 

In Figure \ref{fig:ebv_neb_stell}, we show our measurements of the nebular reddening as a function of the stellar reddening. In the top panels, $\mathrm{E(B-V)_{stellar}}$ is measured using an SMC curve and in the bottom panels using a \citet{Calzetti2000} curve \citep[see equations 8 and 9,][]{Shivaei2020}. The one-to-one relation is shown as a solid line, and the local relation from \citet{Calzetti2000} is shown as a dashed line. The Pearson correlation coefficient for the nebular reddening vs. the stellar reddening indicates a correlation with $\rho=0.15$ and a p-value of $\sim 10^{-3}$. The nebular reddening is on average much higher than the stellar reddening and there is significant scatter between the two. We calculate the reddening ratio, $\mathrm{E(B-V)_{nebular}/E(B-V)_{stellar}}$, only for galaxies where both values of E(B-V)$>0$. The mean of the reddening ratio calculated using the SMC curve (Calzetti curve) for the stellar reddening is 16.7 (14.9) with a standard deviation of 91.3 (129.9). Although these averages are similar, we note that measuring the stellar reddening with the Calzetti curve results in more values with $\mathrm{E(B-V)_{stellar}}<0$. The average reddening ratio values measured at $z\sim 2$ by \citet{Shivaei2020} range from 1.6-2.8, much lower than our average values. The $\mathrm{E(B-V)_{stellar}}$ values in our high redshift sample are much lower ($<0.15$) than their values which range up to $\sim 0.7$. This suggests variations between stellar and nebular dust content that may evolve with redshift.

In the right panels of Figure \ref{fig:ebv_neb_stell}, we show the reddening ratio as a function of the sSFR for individual galaxies with $\mathrm{E(B-V)>0}$. We show the median values as diamonds using the same sSFR bins from Figure \ref{fig:beta_tau}, $-11.8<\rm{log(sSFR/yr^{-1})}\leq -9$ and $-9<\rm{log(sSFR/yr^{-1})}\leq -8$ and $-8< \rm{log(sSFR/yr^{-1})}\leq -6.9$. The highest sSFR bin has significantly higher reddening ratios and large scatter. This is driven by having a range of nebular reddening values with on average very low stellar reddening values for the highest sSFR bin. This suggests that the highly star-forming galaxies in our sample, with high sSFRs, have low diffuse ISM dust but a large range in birth-could dust. Therefore, for high sSFR galaxies at high redshift, there is evidence that on average their dust content seems to be mostly located in their star-forming regions.

We also note that the differences seen between the nebular reddening and stellar reddening could be attributed to variations in the attenuation curve on a galaxy-by-galaxy basis \citep{Markov2025, Shivaei2025}. It is beyond the scope of this paper to measure individual attenuation curves, but this will be explored in future work. 

\section{Summary and Conclusions}\label{sec:summary}
We studied a sample of 602 galaxies in the redshift range $3<z<7$ with 5$\sigma$ measurements of H$\alpha$ and 3$\sigma$ measurements of H$\beta$ emission lines from JADES to investigate the star formation and dust attenuation properties of galaxies at high redshift. Our main results are as follows:

\begin{itemize}

\item We measure SFRs using dust-corrected H$\alpha$ fluxes with a conversion factor that accounts for the sub-solar metallicities that are typical among high redshift galaxies. We construct the star-forming main sequence in two redshift bins, $3\leq z <5$ and $5\leq z \leq 7$. We find that the \citet{Simmonds2025} parameterizations of the SFMS for the median redshifts of our two subsamples are a good match to our data at masses log$\rm{(M_*/M_{\odot})} \gtrsim 8$.

\item We confirm that the correlation between the Balmer decrement and stellar mass is already established out to $z\sim$7. We separate our sample into two redshift bins and find that the Balmer decrement and nebular reddening as a function of stellar mass in galaxies in the redshift range $3\leq z <5$ is consistent with galaxies in the redshift range $5\leq z < 7$. Next, we compare these correlations from our high redshift sample with samples at $z\approx 2$ and $z\approx 0$ and find that they are consistent with each other. This suggests that the Balmer decrement and nebular reddening as a function of stellar mass does not significantly evolve from the local universe to $z\sim$7.

\item We investigate the UV slope as a function of the Balmer optical depth for differing sSFR bins. We find a strong correlation between the UV slope and Balmer optical depth for the lowest sSFR bin, however there is no significant correlation for the highest sSFR bin. This indicates that galaxies in the highest sSFR bin with young stellar ages for our sample have a range of nebular dust attenuation, as measured by the Balmer optical depth, which is on average higher than the UV continuum-inferred dust attenuation, as measured by the UV slope assuming either a Calzetti or an SMC curve. We emphasize that this is evidence that the UV slope should be used with great caution to correct for dust in high redshift galaxies.

\item We compare the nebular reddening to the stellar reddening. The stellar reddening values for our high redshift sample are much lower when compared to studies at lower redshifts. The reddening ratio, $\mathrm{E(B-V)_{nebular}/E(B-V)_{stellar}}$, for our sample is much higher than when compared to galaxies at lower redshift. This suggests variations between stellar and nebular dust content that may evolve with redshift.

\end{itemize}

\section*{Acknowledgments}
This work is based on observations made with the NASA/ESA/CSA JWST. The data were obtained from the Mikulski Archive for Space Telescopes (MAST) at the Space Telescope Science Institute, which is operated by the Association of Universities for Research in Astronomy, Inc., under NASA contract NAS 5–03127 for JWST. These observations are associated with program Nos. 1180, 1181, 1210, 1286, 1895, and 1963. The authors sincerely thank the FRESCO team (PI: Pascal Oesch) and the JEMS team (PI: Christina Williams, Sandro Tacchella, and Michael Maseda) for developing and executing their observing programs. This work is also based (in part) on observations made with the NASA/ESA HST. The data were obtained from the Space Telescope Science Institute, which is operated by the Association of Universities for Research in Astronomy, Inc., under NASA contract NAS 5-26555. Additionally, this work made use of the lux supercomputer at UC Santa Cruz, which is funded by NSF MRI grant AST 1828315.

CW thanks Alice Shapley for providing the median data for Figure \ref{fig:bd_ebv_mass} from \citet{Shapley2022}. We also thank Haoyan Ning for useful discussions.
CW’s research was supported by an appointment to the NASA Postdoctoral Program at the NASA Goddard Space Flight Center, administered by Oak Ridge Associated Universities under contract with NASA.
IS acknowledges funding Atracc{\' i}on de Talento Grant No.2022-T1/TIC-20472 of the Comunidad de Madrid, Spain. IS research is further supported by the European Research Council (ERC) under the European Union’s Horizon 2020 research and innovation programme (Grant agreement No. 101117541, DistantDust).
JW gratefully acknowledges support from the Cosmic Dawn Center through the DAWN Fellowship. The Cosmic Dawn Center (DAWN) is funded by the Danish National Research Foundation under grant No. 140.
AS, AJB, and JC acknowledge funding from the ``FirstGalaxies" Advanced Grant from the European Research Council (ERC) under the European Union’s Horizon 2020 research and innovation programme (Grant agreement No. 789056).
CS acknowledges support from the Science and Technology Facilities Council (STFC), by the ERC through Advanced Grant 695671 “QUENCH”, by the UKRI Frontier Research grant RISEandFALL.
SC acknowledges support by European Union’s HE ERC Starting Grant No. 101040227 - WINGS.
FDE, JS, and RM acknowledge support by the Science and Technology Facilities Council (STFC), by the ERC through Advanced Grant 695671 ``QUENCH'', and by the UKRI Frontier Research grant RISEandFALL.
ECL acknowledges support of an STFC Webb Fellowship (ST/W001438/1).
JMH, BR, and YZ are supported by the JWST/NIRCam Science Team contract to the University of Arizona, NAS5-02015, and JWST Program 3215.
RM also acknowledges funding from a research professorship from the Royal Society.
MP acknowledges grant PID2021-127718NB-I00 funded by the Spanish Ministry of Science and Innovation/State Agency of Research (MICIN/AEI/ 10.13039/501100011033), and the grant RYC2023-044853-I, funded by  MICIU/AEI/10.13039/501100011033 and European Social Fund Plus (FSE+).
ST acknowledges support by the Royal Society Research Grant G125142.
The research of CCW is supported by NOIRLab, which is managed by the Association of Universities for Research in Astronomy (AURA) under a cooperative agreement with the National Science Foundation.

\vspace{5mm}
\facilities{HST, JWST}

\software{\texttt{Prospector} \citep[][]{Johnson2021}, \texttt{python-fsps} \citep[][]{python-fsps}, \texttt{sedpy} \citep[][]{sedpy}, \texttt{fsps} \citep[][]{Conroy2009, Conroy2010}, \texttt{astropy} \citep[][]{astropy:2013, astropy:2018}, \texttt{matplotlib} \citep[][]{matplotlib}, \texttt{dynesty} \citep[][]{dynesty:2020}, \texttt{scipy} \citep[][]{scipy}, \texttt{numpy} \citep[][]{numpy}}




\bibliography{sample631}{}
\bibliographystyle{aasjournal}

\end{document}